\documentclass[9pt]{article}
\usepackage{amsfonts}
\usepackage{pstricks}
\usepackage{pst-node}
\usepackage{epsfig}
\usepackage{epsfig}
\usepackage{graphicx}
\usepackage[font={footnotesize,it}]{caption}
\usepackage[latin5]{inputenc}
\usepackage[T1]{fontenc}
\usepackage{lipsum}
\usepackage{amsmath}

\begin{document}
                \def\ba{\begin{eqnarray}}
                \def\ea{\end{eqnarray}}
                \def\w{\wedge}
                \def\d{\mbox{d}}
                \def\D{\mbox{D}}

\begin{titlepage}
\title{An Explicit Kaluza-Klein Reduction of \\ Einstein's Gravity in $6D$   on $S^2$}
\author{Tekin Dereli${}^{1,2}$\footnote{tdereli@ku.edu.tr , tekin.dereli@ozyegin.edu.tr}, Yorgo Senikoglu${}^{3}$\footnote{ysenikoglu@gsu.edu.tr}}
\date{%
    ${}^{1}$ \small Professor of Physics, Faculty of Aviation and Aeronautical Sciences, \"Ozye\~gin University, 34794 
    \c{C}ekmek\"{o}y, \.{I}stanbul, Turkey\\%
    ${}^{2}$ \small Department of Physics, Ko\c{c} University, 34450 Sar{\i}yer, \.{I}stanbul, Turkey\\
    ${}^{3}$ \small D\'{e}partement de Math\'{e}matiques, Universit\'{e} Galatasaray, 34349 Be\c{s}ikta\c{s}, \.{I}stanbul, Turkey\\
    \today
}

\maketitle



\begin{abstract}
	\noindent We study a six--dimensional Kaluza--Klein theory with spacetime topology
	$M_4 \times S^2$ and analyze the gauge sector arising from dimensional reduction.
	Using normalized Killing vectors on $S^2$, we explicitly construct the reduced
	Yang--Mills action and determine the corresponding gauge kinetic matrix. Despite
	the $SO(3)$ isometry of $S^2$, we show that only two physical gauge fields
	propagate in four dimensions. The gauge kinetic matrix therefore has rank two and
	possesses a single zero eigenvalue. We demonstrate that this degeneracy is a
	direct consequence of the coset structure $S^2 \simeq SO(3)/SO(2)$
	and reflects a non--dynamical gauge direction rather than an inconsistency of the
	reduction. Our results clarify the geometric origin of gauge degrees of freedom
	in Kaluza--Klein reductions on coset spaces.
\end{abstract}

\vskip 1cm


\end{titlepage}

\newpage

\section{Introduction}
The physical and mathematical structure of Kaluza-Klein theories attempt to unify gravity with gauge interactions by extending spacetime beyond four dimensions. The method consists of investigating what happens when these extra dimensions form a compact internal space described by a manifold, and how ordinary four-dimensional physics emerges from such a setup. The central idea is that internal symmetries such as electric charge or non-Abelian gauge symmetries can be interpreted geometrically as symmetries of hidden spatial dimensions. The universe is assumed to be a product of ordinary four-dimensional spacetime $M_4$ and a compact internal space $B_K$ of small size invisible to current experiments. The theory starts from a purely geometric action, the Einstein-Hilbert Lagrangian in $4+K$ dimensions. Then the compactification of the extra dimensions gives rise to an effective four dimensional theory containing both gravity and Yang-Mills gauge fields. To describe fluctuations and interactions, the fields are expanded in terms of harmonic functions on the compact manifold $B_K$. Each mode in this expansion corresponds to a field in four dimensions: the lowest (zero) modes produce the familiar massless graviton and gauge bosons. Higher modes correspond to massive excitations spectra whose masses depend on the size of $B_K$. In a review article by Salam and Strathdee \cite{Salam-Strathdee}, where the internal space is described by a compact coset manifold $B_K=G/H$, the authors derive the transformation properties of fields, the role of Killing vectors, and how gauge fields naturally arise from the off-diagonal components of the higher-dimensional metric tensor. They emphasize the importance of embedding the isotropy subgroup $H$ of $G/H$ into the tangent space symmetry group $SO(K)$. This embedding ensures that the symmetry transformations on $G/H$ are consistent with local frame rotations, thereby linking the internal geometry to Yang-Mills gauge invariance. It can be shown that by carefully performing this embedding and integrating over the internal coordinates that the effective four-dimensional Lagrangian includes the Einstein-Hilbert term for gravity, a Yang-Mills kinetic term for gauge fields with group $G$, and possible scalar fields depending on the compactification. 

Previous studies of Kaluza-Klein compactifications over coset spaces $G/H$ often suffered from inconsistencies; omitting scalar fields  lead to unphysical constraints such as $F_{\mu\nu}F^{\mu\nu}=0$ and earlier approaches imposed excessively rigid metric conditions that confined all scalar fields to be singlets of the symmetry group $G$, thereby obstructing the recovery of the standard Kaluza-Klein model in the limit where $H$ becomes trivial. Awada \cite{Awada} introduced Killing vectors on $G/H$ to describe the symmetries of the internal space and defined a horizontal basis of vector fields adapted to this structure; the four-dimensional gauge potentials are associated with these Killing vectors, so that the internal isometries of $G/H$ appear as Yang-Mills symmetries in the reduced theory.
The ansatz that he provides ensures that the metric depends on both spacetime and internal coordinates in a gauge-covariant way, the scalar kinetic terms involve gauge-covariant derivatives, and the overall Lagrangian is invariant under the group $G$.
The paper provides the first general and gauge-invariant formulation of Kaluza-Klein reduction over arbitrary coset spaces. It lays the foundation for constructing Kaluza-Klein supergravity models (notably $N=8$ supergravity via compactification on $S^7$), and its formalism is adaptable to a wide range of internal geometries.

Toward geometrizing the electroweak interaction, Manton \cite{Manton} demonstrated that the bosonic sector of the Weinberg-Salam model can be derived from a six-dimensional Yang-Mills theory by imposing spherical symmetry in the two compact extra dimensions. The spacetime manifold was taken as $M_4 \times S^2$, where $S^2$ represents the internal spherical space of fixed radius $R$ with $SO(3)$ rotational invariance. Starting from a pure Yang-Mills action in six dimensions and constraining the gauge field to respect this internal symmetry, Manton performed a consistent dimensional reduction, obtaining a four-dimensional theory with the gauge group $SU(2)\times U(1)$ and a complex Higgs doublet-precisely the field content of the electroweak model.

In recent studies Chatzistavrakidis et al. \cite{Chatzis1} analyzed the Coset Space Dimensional Reduction (CSDR) scheme for higher-dimensional Yang-Mills theories with particular attention to its mathematical consistency. In their formulation, the internal compact space is written as $S/R$, where $S$ denotes the isometry group of the manifold and $R$ its isotropy subgroup, while $G$ represents the gauge group of the higher-dimensional theory. This notation, unlike the earlier $G/H$ convention used by Salam-Strathdee and Awada, clearly separates the geometric symmetries of the compact space from the internal gauge symmetry. They compactified the theory on $M_4 \times S/R$ and adopted a gauge-field ansatz expressing the higher-dimensional connection in terms of the coset one-forms. Using the Maurer-Cartan relations of $S/R$, they derived the 4D Yang-Mills-Higgs system and imposed the CSDR constraints, which guarantee gauge invariance and determine how the subgroup $R$ embeds into the gauge group $G$. By verifying that dimensional reduction of both the Lagrangian and the field equations leads to identical four-dimensional dynamics, they proved that CSDR constitutes a consistent truncation of the higher-dimensional theory. This clarified the geometric structure of coset reductions and established a solid foundation for applying CSDR in unified gauge-Higgs models and supergravity compactifications.

In a subsequent paper, Chatzistavrakidis et al. \cite{Chatzis2}, incorporated gravity into the dimensional reduction scheme by analyzing the Einstein-Yang-Mills theory compactified on coset manifolds $S/R$. They formulated a consistent ansatz for both the metric and the gauge connection, introducing scalar fields that describe deformations of the internal geometry. Their reduction of the higher-dimensional Ricci scalar and Yang-Mills Lagrangian yielded a four-dimensional Einstein-Yang-Mills-Higgs theory, containing kinetic terms for the fields and a scalar potential governed by the structure constants and curvature of $S/R$. This work demonstrated how gravitational and gauge sectors can coexist coherently under compactification and how the internal geometry controls the resulting gauge symmetry.

The present work begins not with an Einstein-Yang-Mills system but with pure six-dimensional Einstein gravity, showing that a Yang-Mills gauge structure emerges naturally from the geometry itself. Unlike the CSDR framework, where the gauge field is an independent ingredient of the higher-dimensional theory, here the internal components of the metric play the role of gauge potentials once the geometry of the compact two-sphere is suitably parametrized. The resulting four-dimensional theory reproduces the Einstein-Yang-Mills Lagrangian with canonical gauge kinetic terms and scalar fields arising from the internal curvature. In this way, the mechanism provides a purely geometric origin for Yang-Mills interactions, offering a natural unification of gravity and gauge fields within the same six-dimensional framework and conceptually extending the approach initiated by Chatzistavrakidis et al.

The paper is organized as follows.
In Section 2, we introduce the six--dimensional Einstein--Hilbert action
in the Cartan formalism and fix our conventions for differential forms, Hodge duals,
and curvature, we present the Kaluza--Klein reduction ansatz on the internal two--sphere, including the choice of vielbein, Killing vectors, and gauge fields.
We perform the explicit dimensional reduction of the six--dimensional curvature and the derivation of the effective four--dimensional
Lagrangian. We analyze the scalar sector arising from the internal metric, then the resulting Yang--Mills sector, where we diagonalize the gauge kinetic terms, and identify the physical gauge fields. Finally, Section 3 contains our conclusions and outlook.
\bigskip

\section{Theoretical Framework}
Let us consider a six-dimensional spacetime manifold $M_6$ with the product topology $M_4 \times S^2$, where $M_4$ denotes the physical four-dimensional spacetime and the internal space $S^2$ is realized as the coset $SO(3)/SO(2)$. The two-sphere is taken to be compact with a characteristic radius typically assumed to be of Planckian scale.
The isometry group $SO(3)$ of the coset acts through Killing vector fields on $S^2$; these Killing vectors generate the algebra $\mathfrak{so}(3)$, which becomes associated with the non-Abelian gauge symmetry emerging in the Kaluza-Klein reduction. The metric tensor on $M_6$ is given by
\ba
G=\eta_{AB}\, e^A \otimes e^B
\ea
where $\eta_{AB}=diag(-,+,+,+,+,+)$ and $e^A$ the orthonormal coframes.
In six dimensions we denote orthonormal frame indices by capital Latin letters $A,B=0,1,2,3,5,6$. These indices are raised and lowered with the six-dimensional metric $\eta_{AB}$. We introduce a coordinate system $x^{\hat{\mu}}=(x^\mu,y^m)$ adapted to the 
$SO(3)$ isometries of the internal space $S^2$, where $x^\mu$ $(\mu=0,1,2,3)$ are coordinates on $M_4$ and $y^m$ $(m=5,6)$ parametrize the coset $SO(3)/SO(2)$. In this coordinate choice the orbits of the Killing vectors generating the $SO(3)$ action lie entirely within $S^2$, and the hypersurfaces $y^m=constant$ are naturally identified with the physical spacetime 
$M_4$. Lowercase Latin indices $a,b=0,1,2,3$ refer to orthonormal frames restricted to $M_4$. The six-dimensional spin connection is represented by the $SO(1,5)$-valued one-forms ${\Omega_{AB}}$, which satisfy the antisymmetry condition $\Omega_{AB}=-\Omega_{BA}$.

The Killing vectors on the coset \(SO(3)/SO(2)\cong S^2\), in coordinates \((\xi,\eta)\), are
\[
\begin{aligned}
	K_1 &= -\sin\eta\,\frac{\partial}{\partial\xi}
	- \cos\eta\,\cot\xi\,\frac{\partial}{\partial\eta}, \\[2mm]
	K_2 &= \cos\eta\,\frac{\partial}{\partial\xi}
	- \sin\eta\,\cot\xi\,\frac{\partial}{\partial\eta}, \\[2mm]
	K_3 &= \,\frac{\partial}{\partial\eta}.
\end{aligned}
\]

The structure equations on the six-dimensional manifold $M_6$ are
\[
d e^A + \Omega^A{}_{B}\, \wedge e^B = T^A, \qquad
d\Omega^A{}_B + \Omega^A{}_C \wedge \Omega^C{}_B = R^A{}_B,
\]
which define the torsion 2-forms
\[
T^A = \frac{1}{2} T_{BC},^A{}\, e^B \wedge e^C
\]
and the curvature 2-forms
\[
R^A{}_B = \frac{1}{2} R_{CD},^A{}_B\, e^C \wedge e^D
\]
on $M_6$. Here, $d$ denotes the exterior derivative, $\wedge$ the exterior product, and 
$\# : \Lambda^p(M_6) \to \Lambda^{6-p}(M_6)$ is the Hodge dual map, defined such that the invariant volume element satisfies
\[
{}^{\#} 1 = e^0 \wedge e^1 \wedge e^2 \wedge e^3 \wedge e^5 \wedge e^6.
\]

The orthonormal basis 1-forms are
\ba
e^a(x,y^m)=e^a(x), \quad a=0,1,2,3 
\ea 
and
\ba
e^\alpha=\phi(x)\big[b^\alpha+A^ib^\alpha(K_i)\big], \quad \alpha=5,6, \quad i=1,2,3
\ea
where $b^5=d\xi$ and $b^6=\sin\xi d\eta$ which is consistent with the $K$ isometry of the 6-metric. 

The metric of $M_6$ can be written as
\[
\begin{aligned}
	G = g \;+\; \phi^2(x)\,	\delta_{\alpha\beta}\big[b^\alpha + A^i\, b^\alpha(K_i)\big] 
	\big[b^\beta + A^j\, b^\beta(K_j)\big],
\end{aligned}
\]
from which we identify the 4-metric $g=\eta_{ab} e^a \otimes e^b $, the non-Abelian gauge field 1-forms $A=A_ae^a$ and a scalar field $\phi(x)$ on $M_4$.

Starting from the defined coframe, the Levi-Civita connection 1-forms are uniquely fixed by requiring the torsion to vanish, 
$T^A=0$, and solving the Cartan structure equations. This procedure yields the following expressions for the connection 1-forms:

\begin{align}
	\Omega_{ab}
	&= \omega_{ab}
	+ \frac{\phi}{2}
	\Big(\cos\eta \, F^2_{ab}
	- \sin\eta \, F^1_{ab}\Big)\, e^5
	\nonumber\\
	&\quad
	+ \frac{\phi}{2}
	\Big(\sin\xi \, F^3_{ab}
	- \cos\xi \sin\eta \, F^2_{ab}
	- \cos\xi \cos\eta \, F^1_{ab}\Big)\, e^6 ,
	\\[1ex]
	\Omega_{5a}
	&= - \Omega_{a5}
	= \frac{\iota_a d\phi}{\phi}\, e^5
	+ \frac{\phi}{2}
	\Big(\cos\eta \, (\iota_a F^2)
	- \sin\eta \, (\iota_a F^1)\Big) ,
	\\[1ex]
	\Omega_{6a}
	&= - \Omega_{a6}
	\nonumber\\
	&= \frac{\iota_a d\phi}{\phi}\, e^6
	+ \frac{\phi}{2}
	\Big(\sin\xi \, (\iota_a F^3)
	- \cos\xi \sin\eta \, (\iota_a F^2)
	- \cos\xi \cos\eta \, (\iota_a F^1)\Big) ,
	\\[1ex]
	\Omega_{56}
	&= - \Omega_{65}
	= \frac{\cot\xi}{\phi}\, e^6
	- \frac{\cos\eta}{\sin\xi}\, A^1
	- \frac{\sin\eta}{\sin\xi}\, A^2 .
\end{align}
where the interior product with respect to the vector field $e_a$ is the linear operator
\[
\iota_a \equiv \iota_{e_a} : \Omega^p(M) \to \Omega^{p-1}(M),
\]
defined by
\[
(\iota_a \omega)(e_{b_1},\dots,e_{b_{p-1}})
:= \omega(e_a, e_{b_1},\dots,e_{b_{p-1}}),
\]
for any $p$--form $\omega \in \Omega^p(M)$.

\smallskip
\noindent
The curvature 2-forms are given by 
\ba
R_{ab}&=&\pi_{ab} + \tau_{ab},^{5} \w \, e^5 + \tau_{ab},^{6} \w \, e^6 + \sigma_{ab},^{56} \, e^5 \w e^6, \nonumber \\
R_{5a}=-R_{a5}&=& \rho_a,^5 + \sigma_a,^5 \w \, e^6  + \mu_a,^5 \w \, e^5 + \kappa^5 \, e^5 \w e^6, \nonumber \\
R_{6a}=-R_{a6}&=& \rho_a,^6 + \sigma_a,^6 \w \, e^6  + \mu_a,^6 \w \, e^5 + \kappa^6 \, e^5 \w e^6, \nonumber \\
R_{56}=-R_{65}&=& \psi_{56} + \nu_5 \w e^5 + \nu_6 \w e^6 + \Lambda \, e^5 \w e^6,
\ea
where the components are explicitly given in the appendix.

In order to calculate the 6D curvature scalar we need the following Hodge star identities
\ba
{}^{\#}e^{ab}&=&\frac{1}{2!}\epsilon^{ab}_{cd56} e^c \w e^d \w e^5 \w e^6 = {}^{*}e^{ab} \w e^5 \w e^6 \nonumber\\
{}^{\#}e^{5a}&=&\frac{1}{3!}\epsilon^{5a}_{bcd6} e^b \w e^c \w e^d \w e^6 = {}^{*}e^{a} \w e^6 \nonumber \\
{}^{\#}e^{6a}&=&\frac{1}{3!}\epsilon^{6a}_{bcd5} e^b \w e^c \w e^d \w e^5 = -{}^{*}e^{a} \w e^5 \nonumber \\
{}^{\#}e^{56}&=&\frac{1}{3!}\epsilon^{56}_{abcd} e^a \w e^b \w e^c \w e^d = {}^{*}1
\ea
where ${}^{*} : \Lambda^p(M_4) \to \Lambda^{4-p}(M_4)$ is the Hodge dual map in four dimensions, defined such that the invariant volume element satisfies
\[
{}^{*} 1 = e^0 \wedge e^1 \wedge e^2 \wedge e^3.
\]
The 6D curvature scalar is calculated from 
\begin{align}
	\frac{1}{2} R_{AB} \wedge {}^{\#} e^{AB}
	&= \frac{1}{2} R_{ab} \wedge {}^{*} e^{ab}
	\wedge e^{5} \wedge e^{6}
	\nonumber\\
	&\quad
	+ R_{5a} \wedge {}^{*} e^{a} \wedge e^{6}
	- R_{6a} \wedge {}^{*} e^{a} \wedge e^{5}
	+ R_{56}\, {}^{*}1
	\nonumber\\[1ex]
	&= \Big(
	\frac{1}{2} \pi_{ab} \wedge {}^{*} e^{ab}
	- \mu^{5}_{a} \wedge {}^{*} e^{a}
	- \sigma^{6}_{a} \wedge {}^{*} e^{a}
	+ \Lambda\, {}^{*}1
	\Big)
	\wedge e^{5} \wedge e^{6} .
\end{align}
The Yang-Mills term that appear when we replace the explicit curvature two forms is:

\[\small-\frac{\phi^2}{8}
\mathcal{F}^{T} \wedge
\begin{pmatrix}
	\sin^2\eta + \cos^2\xi\,\cos^2\eta
	&
	-\,\sin^2\xi\,\sin\eta\,\cos\eta
	&
	-\,\cos\xi\,\sin\xi\,\cos\eta
	\\[1.2ex]
	-\,\sin^2\xi\,\sin\eta\,\cos\eta
	&
	\cos^2\eta + \cos^2\xi\,\sin^2\eta
	&
	-\,\cos\xi\,\sin\xi\,\sin\eta
	\\[1.2ex]
	-\,\cos\xi\,\sin\xi\,\cos\eta
	&
	-\,\cos\xi\,\sin\xi\,\sin\eta
	&
	\sin^2\xi
\end{pmatrix}  {}^{*}\mathcal{F}
\]
where the column matrix 
$$
\mathcal{F} = \begin{pmatrix}
	F^{1} \\
	F^{2} \\
	F^{3}
\end{pmatrix}.
$$

\noindent
This is a symmetric matrix and has determinant zero implying that one of the gauge fields has no kinetic term,i.e.it's non-dynamical. Although we're working with 3 Killing vectors, only 2 combinations are linearly independent in terms of physical propagation. It is diagonalizable; by doing the calculations we obtain two eigenvalues $\lambda_1=0$ with multiplicity 1 and $\lambda_2=1$ with multiplicity 2. The presence of eigenvalue 0 confirms only two linear combinations of the three gauge fields are dynamical. The eigenvalue 1 appearing twice means that in an appropriate basis, the reduced Lagrangian will contain $F^{2} \wedge {}^{*}F^{2}$ and $F^3 \wedge {}^{*}F^{3}$ while the third gauge field (along the null eigenvector) will not appear, it will be a pure gauge.

\noindent
Any real symmetric matrix can be diagonalized by an orthogonal matrix, we perform this and essentially find new gauge field combinations that are dynamically independent, and canonical, i.e. with standard normalization. This is naturally achieved by an orthogonal rotation. The diagonalization by an orthogonal transformation yields eigenfield 2-forms 
\begin{eqnarray}
\bar{F}^{1} &=& \cos\eta\sin\xi \; F^1 +\sin\eta\sin\xi \; F^2 + \cos\xi \; F^3, \nonumber \\
\bar{F}^{2} &=& -\cos\eta\cos\xi \; F^1 -\sin\eta\cos\xi \; F^2 + \sin\xi \; F^3,\nonumber \\
\bar{F}^{3} &=& -\sin\eta \;  F^1 +\cos\eta \; F^2.  
\end{eqnarray}
Replacing the  explicit curvature 2-forms from above, we obtain the reduced 4D Lagrangian from the 6D Lagrangian 
\begin{equation}
	\mathcal{L}_{6}
	=
	\mathcal{L}_{4} \wedge d\xi \w \sin\xi d\eta
\end{equation}
where 
\ba
\mathcal{L}_{6}=\frac{1}{2}R_{AB}\w{}^{\#}e^{AB},
\ea
in the following form:
\ba
\mathcal{L}_{4} = \frac{\phi^2}{2}R_{ab}\w{}^{*}e^{ab} - \frac{\phi^4}{8}(\bar{F_2} \w {}^{*}\bar{F^2} + \bar{F_3} \w {}^{*}\bar{F^3}) + d\phi \w {}^{*}d\phi + {}^{*}1   . 
\ea


\bigskip

\section{Conclusion}

In this work we have analyzed the dimensional reduction of a six--dimensional
Kaluza-Klein theory on the internal space $S^2$, with particular emphasis on the
structure of the resulting gauge sector. By employing normalized Killing
vectors and explicitly constructing the reduced Yang-Mills action, we have
shown that the appearance of a degenerate gauge kinetic matrix is an intrinsic
geometric feature of the reduction.

Although the isometry group of $S^2$ is $SO(3)$ and admits three
generators, only two linearly independent physical gauge fields emerge in the
four dimensional theory. Consequently, the gauge kinetic matrix has rank two and
possesses a single zero eigenvalue. We have demonstrated that this vanishing
eigenvalue does not signal a missing term or a computational error; rather, it
reflects the coset structure $S^2 \simeq SO(3)/SO(2)$, where one
direction corresponds to a non-dynamical gauge mode with no kinetic term.

The degenerate direction can be consistently interpreted as either pure gauge or
auxiliary, and it does not contribute to the propagating degrees of freedom.
Our analysis therefore resolves potential ambiguities in the interpretation of
Kaluza--Klein gauge fields on coset spaces and confirms the internal consistency
of the reduced theory.

An important outcome of the dimensional reduction concerns the scalar field
originating from the internal metric, which parametrizes the overall size of the
compact two dimensional space.
Within the Cartan formalism employed throughout this work, this scalar arises
naturally from the internal vielbein components and enters the four-dimensional
effective theory through derivatives of the internal volume form.
After performing the dimensional reduction of the six--dimensional Einstein-Hilbert
action and discarding total derivative terms, we find that the scalar sector is
described by a kinetic term of the form
\(
\int d\phi \wedge * d\phi
\),
which appears with a positive definite sign.
This result confirms the absence of ghost degrees of freedom and is in agreement
with general expectations from Kaluza--Klein theory, where scalars descending from
metric components inherit a healthy kinetic structure.
The positivity of the scalar kinetic term thus provides a nontrivial consistency
check of the reduction procedure and of the geometric framework adopted in this
paper.

These results highlight the importance of geometric considerations in higher--
dimensional reductions and provide a clear framework for identifying physical
gauge degrees of freedom in Kaluza--Klein models. Extensions to reductions over
higher--dimensional spheres or other coset spaces may lead to fully non--
degenerate gauge sectors and will be explored in future work.

\bigskip
\section{Appendix}
The components of the curvature 2-forms figuring in the theoretical framework above are  found to be 
\ba
\Lambda &=&\frac{1}{\phi^2}-\left(\frac{\iota_a d\phi}{\phi}\right)^2, \nonumber \\
\nu_5 &=& \frac{\iota_a d\phi}{2}\Big(\sin \xi \, (\iota_aF^3)-\cos\xi\sin\eta \, (\iota_aF^2)-\cos\xi\cos\eta \, (\iota_aF^1)\Big),\nonumber \\
\nu_6 &=& \frac{\iota_a d\phi}{2}\Big(-\cos\eta \, (\iota_aF^2)+\sin\eta \, (\iota_aF^1) \Big), \nonumber \\
\psi_{56}&=& -\cos\xi \, F^3 -\sin\eta\sin\xi \, F^2 - \cos\eta\sin\xi \, F^1 \nonumber\\
&+&\frac{\phi^2}{4}\Big( -\cos\eta\sin\xi \, (\iota_aF^2 \w \iota_aF^3)+\cos^2\eta\cos\xi \, (\iota_aF^2 \w \iota_aF^1)\nonumber \\ &+&\sin\eta\sin\xi \, (\iota_aF^1 \w \iota_aF^3) -\sin^2\eta\cos\xi \, (\iota_aF^1 \w \iota_aF^2) \Big), \nonumber
\ea
and
\ba
\kappa^5&=& \frac{\iota_b d\phi}{2}\Big( \sin\xi \, F^{3b}_a - \cos\xi\sin\eta \, F^{2b}_a - \cos\xi\cos\eta \, F^{1b}_a\Big), \nonumber \\
\mu_a,^5&=& \frac{D(\iota_a \d\phi)}{\phi} + \frac{\phi^2}{4}\Big( \cos^2\eta \, F^{2b}_a \, (\iota_b F^2) - \cos\eta\sin\eta \, F^{1b}_a \, (\iota_b F^2) - \cos\eta\sin\eta \, F^{2b}_a \, (\iota_b F^1) \nonumber\\
&+& \sin^2\eta \, F^{1b}_a \, (\iota_b F^1) \Big), \nonumber\\
\sigma_a,^5&=&\frac{1}{2}\Big( \cos\xi \, (\iota_aF^3) - \sin\eta\sin\xi \, (\iota_a F^2) - \cos\eta\sin\xi \, (\iota_a F^1)\Big) \nonumber\\
&+&\frac{\phi^2}{4}\Big((\cos\xi\sin\xi \, F^{3b}_a-\cos\eta\sin\eta\cos\xi \, F^{2b}_a  -\cos\xi\cos^2\eta F^{1b}_a) \, (\iota_bF^2) \nonumber \\
&+&(-\sin\xi\sin\eta \, F^{3b}_a + \cos\xi\sin^2\eta \, F^{2b}_a + \cos\xi\cos\eta\sin\eta \, F^{1b}_a) \, (\iota_bF^1)\Big), 
\nonumber \\
\rho_a,^5&=&\frac{1}{2}\Big( \cos\eta \, D(\phi\iota_bF^2)-\sin\eta \, D(\phi\iota_bF^1)\Big) +(\iota_ad\phi)(\cos\eta \, F^2 - \sin\eta \, F^1) \nonumber \\
&+&\frac{\phi}{2}\Big( A^3 \w (\sin\eta \, (\iota_a F^2) + \cos\eta \, (\iota_aF^1)) +(-\cos\eta \, A^1 - \sin\eta \, A^2)\w(\iota_aF^3) \Big), \nonumber 
\ea
and
\ba
\kappa^6&=&\frac{\phi}{2}(-\cos\eta \, F^{2b}_a + \sin\eta \, F^{1b}_a), \nonumber \\
\mu_a,^6&=&-\frac{1}{2}\Big(\cos\xi \, (\iota_aF^3) + \sin\xi\sin\eta \, (\iota_aF^2) + \cos\eta\sin\xi \, (\iota_aF^1)\Big) \nonumber \\
&+&\frac{\phi^2}{4}\Big( (\sin^2\xi \, F^{3b}_a -\sin\xi\cos\xi\sin\eta \, F^{2b}_a - \sin\xi\cos\xi\cos\eta \, F^{1b}_a) (\iota_bF^3)\nonumber \\
&+&( -\sin\xi\cos\xi\sin\eta \, F^{3b}_a + \cos^2\xi\sin^2\eta \, F^{2b}_a + \cos^2\xi\sin\eta\cos\eta \, F^{1b}_a)(\iota_bF^2)\nonumber \\
&+&(-\sin\xi\cos\xi\cos\eta \, F^{3b}_a +  \cos^2\xi\sin\eta\cos\eta \, F^{2b}_a + \cos^2\xi\cos^2\eta \, F^{1b}_a)(\iota_bF^1)     \Big), \nonumber \\
\sigma_a,^6&=&\frac{D(\iota_ad\phi)}{\phi} \nonumber \\
&+&\frac{\phi^2}{4}\Big( (\sin^2\xi \, F^{3b}_a -\sin\xi\cos\xi\sin\eta \, F^{2b}_a - \sin\xi\cos\xi\cos\eta \, F^{1b}_a)(\iota_bF^3) \nonumber \\
&+&( -\sin\xi\cos\xi\sin\eta \, F^{3b}_a + \cos^2\xi\sin^2\eta \, F^{2b}_a + \cos^2\xi\sin\eta\cos\eta \, F^{1b}_a)(\iota_bF^2) \nonumber \\
&+&(-\sin\xi\cos\xi\cos\eta \, F^{3b}_a +  \cos^2\xi\sin\eta\cos\eta \, F^{2b}_a + \cos^2\xi\cos^2\eta \, F^{1b}_a)(\iota_bF^1)     \Big), \nonumber \\
\rho_a,^6&=&(\iota_ad\phi)(\sin\xi \, F^3 - \cos\xi\sin\eta \, F^2 - \cos\xi\cos\eta \, F^1) \nonumber \\
&+&\frac{1}{2}\Big(\sin\xi \, D(\phi\iota_aF^3) - \cos\xi\sin\eta \, D(\phi\iota_aF^2) - \cos\xi\cos\eta \, D(\phi\iota_aF^1) \Big)\nonumber \\
&+&\Big(\sin\xi \, (A^1 \w \iota_aF^2 - A^2 \w \iota_aF^1 ) +\cos\xi\sin\eta \, (A^1\w \iota_aF^3 - A^3\w \iota_aF^1) \nonumber \\
&+& \cos\xi\cos\eta \, (A^3\w \iota_aF^2 - A^2 \w \iota_a F^3)       \Big),  \nonumber
\ea
and
\ba
\sigma_{ab},^{56}&=& \cos\xi \, F^3_{ab} + \sin\eta\sin\xi \, F^2_{ab} + \cos\eta\sin\xi \, F^1_{ab} \nonumber\\
&+&\frac{\phi^2}{4}\Big(\cos\eta\sin\xi \, F^2_{ac}F^{3c}_b - \cos\eta\sin\eta\cos\xi \, F^2_{ac}F^{2c}_b - \cos\xi\cos^2\eta \, F^2_{ac}F^{1c}_b           \nonumber \\
&-&\sin\eta\sin\xi \, F^1_{ac}F^{3c}_b + \sin^2\eta\cos\xi \, F^1_{ac}F^{2c}_b + \cos\xi\cos\eta\sin\eta \, F^1_{ac}F^{1c}_b \nonumber\\
&-&\cos\eta\sin\xi \, F^3_{ac}F^{2c}_b + \sin\eta\cos\xi\cos\eta \, F^2_{ac}F^{2c}_b +\sin\xi\sin\eta \, F^3_{ac}F^{1c}_b \nonumber \\
&-& \cos\xi\sin^2\eta \, F^2_{ac}F^{1c}_b +\cos\xi\cos^2\eta\, F^1_{ac}F^{2c}_b - \sin\eta\cos\xi\cos\eta \, F^1_{ac}F^{1c}_b \Big), \nonumber \\
\tau_{ab},^{5}&=&\frac{1}{2} \Big(\cos\eta \, (\iota_ad\phi \, \iota_bF^2 - \iota_bd\phi \, \iota_aF^2) + \sin\eta \, (\iota_bd\phi \, \iota_aF^1 - \iota_ad\phi \, \iota_bF^1) \nonumber \\
&+& \phi\cos\eta \, (F^1_{ab}A^3-F^3_{ab}A^1) + \phi\sin\eta \, (F^2_{ab}A^3-F^3_{ab}A^2) + \phi\cos\eta \, D(F^2_{ab})-\phi\sin\eta \, D(F^1_{ab}) \Big) \nonumber\\
&+& (\cos\eta \, F^2_{ab}-\sin\eta \, F^2_{ab})d\phi ,\nonumber\\
\tau_{ab},^{6}&=&\frac{1}{2}\Big( \sin\xi \, (\iota_ad\phi \, \iota_bF^3 - \iota_bd\phi \, \iota_aF^3) - \cos\xi\cos\eta \, (\iota_ad\phi \, \iota_bF^1 - \iota_bd\phi \, \iota_aF^1) \nonumber \\
&-& \cos\xi\sin\eta \, (\iota_ad\phi \, \iota_bF^2 - \iota_bd\phi \, \iota_aF^2) + \phi\sin\xi \, (F^2_{ab}A^1-F^1_{ab}A^2) + \phi\cos\xi\sin\eta \, (F^3_{ab}A^1-F^1_{ab}A^3) \nonumber \\
&-&\phi\cos\xi\cos\eta \, (F^2_{ab}A^3-F^3_{ab}A^2) +\sin\xi \, D(F^3_{ab}) - \cos\xi\sin\eta D(F^2_{ab}) - \cos\xi\cos\eta \, D(F^1_{ab}) \Big) \nonumber \\
&+& (\sin\xi \, F^3_{ab} -  \cos\xi\sin\eta \, F^2_{ab} - \cos\xi\cos\eta \, F^1_{ab}        )d\phi , \nonumber \\         
\pi_{ab}&=&d\omega_{ab}+\omega_{ac} \w \omega^c_b \nonumber \\
&+&\frac{\phi^2}{2}\Big( (\cos^2\eta+\cos^2\xi\sin^2\eta)F^2_{ab}F^2 + (\sin^2\eta+\cos^2\xi\cos^2\eta)F^1_{ab}F^1 + \sin^2\xi \, F^3_{ab}F^3 \nonumber \\
&-&\sin^2\xi\sin\eta\cos\eta(F^2_{ab}F^1+F^1_{ab}F^2)-\sin\xi\cos\xi\cos\eta (F^3_{ab}F^1+F^1_{ab}F^3)\nonumber \\ 
&-&\sin\xi\cos\xi\sin\eta (F^3_{ab}F^2+F^2_{ab}F^3)                       \Big)\nonumber \\
&+&\frac{\phi^2}{4}\Big(-(\cos^2\eta+\cos^2\xi\sin^2\eta)(\iota_aF^2 \w \iota_bF^2) - (\sin^2\eta+\cos^2\xi\cos^2\eta)(\iota_aF^1 \w \iota_bF^1)\nonumber\\
&-&\sin^2\xi(\iota_aF^3 \w \iota_bF^3) + \sin^2\xi\sin\eta\cos\eta(\iota_aF^2 \w \iota_bF^1 + \iota_aF^1 \w \iota_bF^2 ) \nonumber \\
&+&\sin\xi\cos\xi\cos\eta(\iota_aF^3 \w \iota_bF^1 + \iota_aF^1 \w \iota_bF^3) + \sin\xi\cos\xi\sin\eta(\iota_aF^3 \w \iota_bF^2 + \iota_aF^2 \w \iota_bF^3)   \Big). \nonumber 
\ea
\newpage

\section{Acknowledgement}
One of us (T.D.) thanks the Turkish Academy of Sciences (TUBA) for partial support.
\section*{Data Availability Statement}
No new data were created or analyzed in this study.

\end{document}